\documentclass{amsart}

\newtheorem{theorem}{Theorem}[section]

\theoremstyle{definition}

\theoremstyle{remark}

\usepackage{graphicx}
\usepackage{color}
\usepackage{amsmath}
\usepackage{amsfonts}
\usepackage{amssymb}
\newcommand{\be}{\begin{equation}}
\newcommand{\ee}{\end{equation}}

\newcommand{\dz}{\wedge}

\newcommand{\ba}{\begin{array}}

\newcommand{\ea}{\end{array}}

\newcommand{\beq}{\begin{eqnarray}}

\newcommand{\eeq}{\end{eqnarray}}

\newtheorem{lm}{lemma}

\newtheorem{thee}{theorem}

\newtheorem{proo}{proposition}

\newtheorem{co}{corollary}

\newtheorem{rem}{remark}

\newtheorem{deff}{definition}

\newcommand{\bd}{\begin{deff}}

\newcommand{\ed}{\end{deff}}

\newcommand{\bl}{\begin{lm}}

\newcommand{\el}{\end{lm}}

\newcommand{\bp}{\begin{proo}}

\newcommand{\ep}{\end{proo}}

\newcommand{\bt}{\begin{thee}}

\newcommand{\et}{\end{thee}}

\newcommand{\bc}{\begin{co}}

\newcommand{\ec}{\end{co}}

\newcommand{\brm}{\begin{rem}}

\newcommand{\erm}{\end{rem}}

\newcommand{\der}{{\rm d}}

\usepackage{t1enc}

\newcommand{\newc}{\newcommand}

\let\ccdot\cdot
\def\cdot{\hbox to 2.5pt{\hss$\ccdot$\hss}}

\newc{\aR}{\mbox{\boldmath{$ R$}}}
\newc{\aS}{\mbox{\boldmath{$ S$}}}
\newc{\aT}{\mbox{\boldmath{$ T$}}}
\newc{\aW}{\mbox{\boldmath{$ W$}}}

\newc{\aK}{\mbox{\boldmath{$ K$}}}
\newc{\aL}{\mbox{\boldmath{$ L$}}}

\newcommand{\bbC}{\mathbb{C}}

\usepackage{amssymb}
\usepackage{amscd}

\newcommand{\hook}{\raisebox{-0.35ex}{\makebox[0.6em][r]
{\scriptsize $-$}}\hspace{-0.15em}\raisebox{0.25ex}{\makebox[0.4em][l]{\tiny
 $|$}}}

\newc{\obstrn}[2]{B^{#1}_{#2}}

\newcommand{\rpl}                         
{\mbox{$
\begin{picture}(12.7,8)(-.5,-1)
\put(0,0.2){$+$}
\put(4.2,2.8){\oval(8,8)[r]}
\end{picture}$}}

\newcommand{\lpl}                         
{\mbox{$
\begin{picture}(12.7,8)(-.5,-1)
\put(2,0.2){$+$}
\put(6.2,2.8){\oval(8,8)[l]}
\end{picture}$}}

\usepackage{ifthen}

\newcommand{\bbR}{\mathbb{R}}
\newcommand{\bbS}{\mathbb{S}}

\newc{\tensor}[1]{#1}
\newc{\Mvariable}[1]{\mbox{#1}}
\newc{\down}[1]{{}_{#1}}
\newc{\up}[1]{{}^{#1}}

\newc{\JulyStrut}{\rule{0mm}{6mm}}
\newc{\midtenPan}{\mbox{\sf S}}
\newc{\midten}{\mbox{\sf T}}
\newc{\midtenEi}{\mbox{\sf U}}
\newc{\ATen}{\mbox{\sf E}}
\newc{\BTen}{\mbox{\sf F}}
\newc{\CTen}{\mbox{\sf G}}

\def\sideremark#1{\ifvmode\leavevmode\fi\vadjust{\vbox to0pt{\vss
 \hbox to 0pt{\hskip\hsize\hskip1em
 \vbox{\hsize3cm\tiny\raggedright\pretolerance10000
 \noindent #1\hfill}\hss}\vbox to8pt{\vfil}\vss}}}%

\numberwithin{equation}{section}

\newcommand{\bma}{\begin{pmatrix}}
\newcommand{\ema}{\end{pmatrix}}

\newcounter{romenumi}
\newcommand{\labelromenumi}{(\roman{romenumi})}

\begin{document}
\title{On Penrose's `before the big bang' ideas}

\author{C. Denson Hill} \address{Department of Mathematics, Stony
  Brook University, Stony Brook, N.Y. 11794, USA}
\email{dhill@math.sunysb.edu}
\author{Pawe\l~ Nurowski} 
\address{Instytut Fizyki Teoretycznej,
Uniwersytet Warszawski, ul. Hoza 69, 00-681 Warszawa, Poland}
\email{nurowski@fuw.edu.pl} 
\thanks{This research was supported by the KBN grant 1 P03B 07529}
\date{\today}

\begin{abstract}
We point out that algebraically special Einstein fields with
twisting rays exhibit the basic properties of conformal Universes
considered recently by Roger Penrose.   
\vskip5pt\centerline{\small\textbf{MSC classification}: 83C15, 83C20, 83C30,
  83F05}\vskip15pt
\end{abstract}
\maketitle
According to Penrose \cite{pen}, there might have been a 
\emph{pre-big-bang era}, in which the Universe was equipped 
with a  
\emph{conformal} Lorentzian structure, rather than with the \emph{full}
Lorentzian \emph{metric} structure. This conformal structure was conformally
flat, and pretty much the same as the conformal structure our Universe
will have at the \emph{very late stage} of its evolution. 

Penrose argues that 
the observed \emph{positive sign} of the cosmological constant,
$\Lambda>0$, forces our Universe to last forever. It will last 
long enough that all the
massive particles will mannage to disintegrate, either by finding 
their antimatter counterparts with which to annihilate, or because of
their finite half life. This will produce only \emph{massless}
particles, such as photons, which after all the matter has been
disintegrated, will be the only 
content of the late Universe, except for massive \emph{black holes}. These
massive black holes will be the remnants of the galaxies, and
perhaps, of very massive stars. 

Since the Universe will last forever, and since it will be expanding,
it will generally \emph{cool down}, trying to reach the \emph{absolute
  zero} temperature at its final state. Thus, there will be a time in 
its evolution such that the temperature of the Universe will be lower
than the temperature of even the 
\emph{most massive} black holes, which at this stage will hide the
only \emph{mass} of the Universe. This will create a thermodynamic instability forcing
\emph{all} the black holes to evaporate by radiating massless
particles\footnote{Penrose \emph{assumes} that the black holes lose
  information about the properties of the masses hidden in them, and
  that in the process of evaporation only massless particles are
  being radiated.}. After these evaporations the dying Universe will be
\emph{totally} filled by \emph{massless particles}. 

Penrose argues that massless
particles, whatever they are, have no way of defining
clocks' ticks. This leads to the conclusion that the Universe at its
late age, being filled with \emph{only} massless observers, will
ultimately lose the information about its conformal factor. Thus it will
become similar to the Universe in the `pre-big-bang' era that
preceeded its creation. This late state of the Universe
Penrose likes \emph{to interprete} as the `pre-big-bang' era of the
\emph{new} Universe. Let us adopt this point of view.

It is a well known fact that during the \emph{big bang} (such as in the Friedmann-Lemaitre-Robertson-Walker
models) the only metric
\emph{singularity}  is in the \emph{vanishing of the conformal factor}, leaving the conformally rescaled metric perfectly regular 
(actually \emph{conformally flat}). Another well known fact is that
the \emph{field equations} for the \emph{massless} particles are \emph{conformally
invariant}. This shows that 
the massless particles (the observers) of the dying conformal
Universe, or as we
interpret it now, of the `pre-big-bang'
era of the \emph{new} conformal Universe, \emph{will not feel} the big
bang singularity. They will happily \emph{pass through it} from the
dying conformal Universe to
another conformally flat manifold\footnote{It should be noted that
  this passage is only possible \emph{conformally}. In the \emph{full} Lorentzian
  \emph{metric} of the old Universe, the process of expansion will last
  \emph{infinitely long}. But after conformal rescaling, this proces
  takes \emph{finite time}, and enables us to speculate what will be after.}. Penrose interprets this conformal
object as an
`after big bang' conformal era of the new Universe. It will eventually
acquire a \emph{new conformal factor}, and possibly some distortion
(meaning non-zero Weyl), 
promoting this conformal remnant of the old Universe to  
a \emph{new} Lorentzian \emph{metric} Universe.   \\  

Our \emph{exact solutions of Einstein's equations}, discussed below,
exhibit the main features of Penrose's `before the
big bang' argument. Although these solutions form a very \emph{thin
  set} in all the 
possible Einstein metrics, and although they were
obtained on \emph{purely mathematical grounds} by the mere assumption that the
corresponding spacetimes admit a \emph{twisting congruence of null and shearfree
  geodesics}, it is a \emph{remarkable coincidence} that the 
\emph{pure mathematics} of Einstein's equations, imposed on such
spacetimes, forces the solutions to fit to Penrose's ideas. \\

Let us discuss the mathematics of our solutions first. After we do it, we will
indicate the parallels between our solutions and Penrose's Universes.\\

About twenty years ago we proved \cite{ln,nurphd} the following 
\begin{theorem}
Let $({\mathcal M},g)$ be a 4-dimensional Lorentzian spacetime which satisfies 
the Einstein equations
\be
R_{\mu\nu}=\Lambda g_{\mu\nu}+\Phi k_\mu k_\nu,\quad\quad\Lambda={\rm const},\label{ee}
\ee
with $k_\mu$ being a null vector tangent to 
a \emph{twisting} congruence of
\emph{null} and \emph{shearfree geodesics}. Then its metric $g$ \emph{factorizes} as
\be
g=\Omega^{-2} \hat{g},\quad\quad \Omega=\cos(\tfrac{r}{2}),\label{met}\ee
where $\hat{g}$ is \emph{periodic} in terms of the null
cooordinate $r$ along $k=\partial_r$.
\end{theorem}
 More explicitly, (see 
\cite{ln,nurphd} and, especially, \cite{lhn} for details) we showed that if $g$
satisfies (\ref{ee}), then $\mathcal M$ is a \emph{circle bundle}
$\bbS^1\to\mathcal M\to M$ over a 3-dimensional strictly pseudoconvex 
CR manifold $(M,[(\lambda,\mu)])$, and that 
\be \hat{g}=p^2[\mu\bar{\mu}+\lambda(\der
r+W\mu+\bar{W}\bar{\mu}+H\lambda)],\label{cmet}\ee
with
\be
W=i a {\rm e}^{-i r}+b,\quad H=\frac{n}{p^4}{\rm e}^{2i
  r}+\frac{\bar{n}}{p^4}{\rm e}^{-2i r}+q{\rm e}^{i r}+\bar{q}{\rm
  e}^{-i r}+h.\label{w}\ee
Here $\lambda$ (real) and $\mu$ (complex) are 1-forms on $\mathcal M$ such
that $k\hook\lambda=k\hook\mu=0$, $k\hook\der\lambda=k\hook\der\mu=0$, 
$\lambda\dz\mu\dz\bar{\mu}\neq 0$ and, as a result of the Einstein equations (\ref{ee}), the 
functions $a,b,n,q$ (complex) and $p,h$ (real) are \emph{independent}
of the null coordinate $r$: $a_r=b_r=n_r=q_r=p_r=h_r=0$. A part 
of the Einstein
equations in (\ref{ee}) can be explicitly integrated, obtaining:
\begin{eqnarray}
a&=&c+2\partial\log p\nonumber\\
b&=&ic+2i\partial\log p\nonumber\\
q&=&\frac{3n+\bar{n}}{p^4}+\frac23\Lambda p^2+\frac{2\partial p\bar{\partial}p-p(\partial\bar{\partial}p+\bar{\partial}\partial
  p)}{2p^2}-\frac{i}{2}\partial_0\log
  p-\bar{\partial}c\label{q}\\
h&=&3\frac{n+\bar{n}}{p^4}+2\Lambda p^2+\frac{2\partial
  p\bar{\partial}p-p(\partial\bar{\partial}p+\bar{\partial}\partial
  p)}{p^2}-\bar{\partial}c-\partial\bar{c}.\nonumber
\end{eqnarray}
Here the $r$-independent complex function $c$ is defined via
\begin{eqnarray}
&&\der\mu=0,\quad\quad\quad\der\bar{\mu}=0,\nonumber\\
\der\lambda&=&i\mu\dz\bar{\mu}+(c\mu+\bar{c}\bar{\mu})\dz\lambda,\label{c}
\end{eqnarray}
and the operators $(\partial_0,\partial,\bar{\partial})$ are 
vector fields on $M$, which
are algebraic dual to the coframe $(\lambda,\mu,\bar{\mu})$ on
the CR manifold $M$. Note that the function $c$ is \emph{defined
uniquely}, once a CR manifold $(M,[(\lambda,\mu)])$ has been chosen, and thus
is considered as a \emph{known} function in the process of solving
(\ref{ee}). 

The remaining Einstein equations in (\ref{ee})
reduce to a system of \emph{two} PDEs on $M$, for the functions $n$ and
$p$, which are the only \emph{unknowns}. These PDEs are:
\begin{eqnarray}
&&\partial n+3cn=0,\label{m}\\
&&[~\partial\bar{\partial}+\bar{\partial}\partial
   +\bar{c}\partial+c\bar{\partial}+\tfrac12c\bar{c}+\tfrac34(\partial\bar{c}+\bar{\partial}
    c)~]p=\frac{n+\bar{n}}{p^3}+\tfrac23\Lambda p^3.\label{p}
\end{eqnarray}    
These are the only equations which need to be solved in order to make
$g$ satisfy (\ref{ee}). Once these equations are solved, the Einstein metric $g$
has an energy momentum tensor describing the `pure radiation' of a mixture of
\emph{massless} particles, moving with the speed of light along the null
direction $k$, in a spacetime with cosmological constant
$\Lambda$. The spacetime is algebraically special; the 
Weyl spin coefficients $\Psi_0,\Psi_1,\Psi_2$ being $\Psi_0=\Psi_1=0$, 
$\Psi_2=\frac{(1+{\rm e}^{ir})^3}{2p^6}~n$.

Note that at $r=\pm\pi$, where the \emph{conformal factor}
$\Omega$ for $g$ becomes \emph{zero}, the Weyl coefficient $\Psi_2$ \emph{vanishes},  $$\Psi_2(\pm\pi)=0.$$ Although the formulae for $\Psi_3$ and $\Psi_4$
are quite complicated, they also share this property, i.e.
$$\Psi_3(\pm\pi)=\Psi_4(\pm\pi)=0.$$

The above quoted result enables us to interpret the $r=\pm\pi$
hypersurfaces as the respective scris ${\mathcal I}^\pm$ of the spacetime
$({\mathcal M},g)$. The Weyl tensor is conformally flat there: $\Psi_\mu(\pm\pi)=0$ for all $\mu=0,1,2,3,4$.

Since the conformally rescaled spacetime $({\mathcal M},\hat{g})$ is 
\emph{periodic} and \emph{regular} in $r$, it gives a
full-Einstein-theory realization of Penrose's idea \cite{pen}
that there was a `before the big bang era' of the Universe. \\

In this context the following remarks are in order:
\begin{itemize}
\item To be in accordance with current observations, we  
concentrate on those solutions (\ref{met})-(\ref{cmet})
  of the Einstein equations (\ref{ee}) which have \emph{positive}
  cosmological constant $\Lambda>0$. 
\item It is known (\cite{ss}, p. 353) that every solution of Einstein's
  equations (\ref{ee}) with $\Lambda>0$ has \emph{spacelike} scris ${\mathcal
    I}^\pm$. Thus, restricting to $\Lambda>0$, we know that our
  scris ${\mathcal I}^\pm$ at $r=\pm\pi$ are \emph{spacelike}.
\item Moreover the scris are \emph{conformally flat}, and therefore can be
  identified with the respective surfaces of the \emph{big bang}
  ($r=-\pi$) and the surface of the \emph{conformal infinity in the
    future} 
  ($r=\pi$).
\item The Universes corresponding to our solutions are either
  \emph{empty} ($\Phi=0$), or are filled with a dust of
  \emph{massless} particles ($\Phi> 0$) moving with the speed of light along $k$.
\item Since going from the `big bang' to the `infinity in the future'
  corresponds to a passage from $r=-\pi$ to $r=\pi$, and since the
  conformal metric $\hat{g}$ is \emph{periodic} in $r$, we see that our
  \emph{conformal} solutions are repetitive in the $r$ variable. 
\item Thus our solutions give \emph{conformal Universes which   
  periodically reproduce themselves, and smoothly pass through the `big
  bang' and the `future infinity'}.
\end{itemize}

To be more explicit we discuss the following example\footnote{Note
that among our solutions there are many interesting, well known, solutions of
Einstein's field equations. Actually our metrics include all
\emph{algebraically special} vacuums and the aligned pure radiation
gravitational fields. Thus, our solutions include for example the celebrated 
\emph{rotating black hole} solution of Kerr. This solution, however, 
is beyond the class of solutions relevant for Penrose's
ideas since it has $\Lambda=0$ (and $\Phi=0$).} which is  
a solution to our equations (\ref{m})-(\ref{p}). \\

We choose the CR manifold $(M,[(\lambda,\mu)])$
to be the Heisenberg group CR manifold. This may be represented by the
1-forms $\lambda=\der u+\frac{i}{2}(\bar{z}\der z-z\der\bar{z})$ and 
$\mu=\der z$. Here $(u,z,\bar{z})$ are the standard coordinates on the 
Heisenberg
group ($u$ is real, $z$ is complex).

Obviously $\der\mu=0$ and
$\der\lambda=i\mu\dz\bar{\mu}$, so that the function $c$ in (\ref{c})
is $c\equiv 0$. This immediately leads to a solution for
(\ref{m})-(\ref{p}). Indeed: take $n=const\in\bbC$ and $p=1$, then
equation (\ref{m}) is automatically satisfied, and equation (\ref{p})
gives $\Lambda=-\frac32(n+\bar{n})$. Thus we take
$n=-\tfrac13\Lambda+im=const$, $\Lambda,m\in\bbR$. 
This leads to the \emph{conformal metric} 
$$
\hat{g}=2\cos^2(\tfrac{r}{2})g=2\der z\der\bar{z}+2\lambda\Big(\der
r-2\big(2m(1+\cos r)\sin r+\frac{\Lambda}{3}(2\cos r+\cos
2r)\big)\lambda\Big),$$
with $\lambda=\der u+\frac{i}{2}(\bar{z}\der z-z\der\bar{z})$, 
and the \emph{physical metric} $g$ satisfying all the equations
(\ref{ee}). Actually, the physical metric satisfies more: It is a
solution to the Einstein equations $R_{\mu\nu}=\Lambda g_{\mu\nu}$,
thus $\Phi=0$. Its Weyl tensor has $\Psi_0=\Psi_1=\Psi_3=\Psi_4=0$
everywhere, with the only nonvanishing Weyl coefficient
$\Psi_2=-\tfrac13(1+{\rm e}^{ir})^3(\Lambda-3im).$ This means that the
metric corresponding to this solution is of Petrov type $D$
everywhere, except along the scris, $r=\pm\pi$, where it is conformally
flat\footnote{This solution $g$ is the classical Taub-NUT solution with
  cosmological constant\cite{ca,pd}. It includes the Taub-NUT
  ($\Lambda=0$) solution \cite{nut} as
  a special case. When $m=0$ the solution is a vacuum metric with a
  cosmological constant, which has two Robinson congruences \cite{pr} as two
  distinct principal null dierctions.}. Restricting
to $\Lambda>0$ we get a 2-parameter family of solutions with spacelike scris,
which has a periodic conformal metric $\hat{g}$. This, in addition to
being periodic in $r$, is 
regular everywhere on any hypersurface transversal to $k$.\\

As a more complicated example we take a CR structure parameterized by
coordinates $(x,y,u)$ and represented by
forms 
$$
\lambda=-\frac{2}{f'(y)}\Big(\der u+f(y)\der x\Big),\quad\quad\mu=\der
x+i\der y.$$
It has $c=\frac{if''(y)}{2f'(y)}$. Then assuming that $p=p(y)$ and
$n=n(y)$, we immediately get the following solution for equation
(\ref{m}): 
$$n=f'(y)^3m,\quad\quad\quad m={\rm const}\in\bbC.$$
Having this,  
the only remaining Einstein equation to be solved is (\ref{p}). It is equivalent
to an ODE:
\begin{eqnarray*}
\tfrac94 pf'f'''+3p'f'f''-3p{f''}^2-3p''{f'}^2+4\Lambda p^3 {f'}^2+6(m+\bar{m}){f'}^5p^{-3}=0,
\end{eqnarray*}
for the functions $p=p(y)$ and $f=f(y)$. Since this is a \emph{single} ODE for \emph{two}
real functions of one real variable $y$, one can use one of these functions to arrange 
the energy $\Phi$ of the corresponding pure radiation to be nonegative for
positive $\Lambda$.\\

We believe that many more solutions with appealing
physical properties may be found in our class, the main reason being
that the class consists of \emph{all} (known and unknown) algebraically special
solutions with twisting rays. \\

We close the paper with the following mathematical comment.

It is interesting to give an interpretation to the only
nontrivial Einstein equation (\ref{p}). If one considers the metric 
$\tilde{g}=\sec^2(\tfrac{r}{2})\hat{g}$,
with $\hat{g}$ as in (\ref{cmet}), and functions $W,H$ as in
(\ref{w})-(\ref{q}), then the equation (\ref{p}) is the Yamabe
equation (see e.g. \cite{ss}, p. 332)
saying that the rescaled metric $g=p^2\tilde g$ has constant Ricci
scalar $R=4\Lambda$. 

\end{document}